\documentclass[12pt]{article}
\usepackage{epsfig}

\usepackage{amssymb}
\usepackage{amsfonts}

\usepackage{color}
 
\def\be{\begin{equation}}
\def\ee{\end{equation}}
\def\ba{\begin{array}{c}}
\def\ea{\end{array}}

\newcommand{\bea}{\begin{eqnarray}}
\newcommand{\eea}{\end{eqnarray}}

\newcommand{\kt}{\rangle}

\begin{document}

\begin{center}

{\Large \bf

Triple exceptional point with unitary paths of unfolding
in a three-site fermionic Swanson-like model

}

\end{center}

\vspace{0.4cm}

\begin{center}

 {\bf Bijan Bagchi}$^{a}$,
  {\bf
Aritra Ghosh}$^{b,}$\footnote{present address:
School of Physics and Astronomy, Rochester Institute of Technology,
Rochester, New York 14623, USA}
and
  {\bf Miloslav Znojil}$^{c,d,e,}$\footnote{corresponding author,
{e-mail: znojil@ujf.cas.cz}}

\end{center}

$^{a}$
Department of Applied Mathematics,
University of Calcutta,
Kolkata 700009,
India



$^{b}$
School of Basic Sciences, Indian Institute of Technology, Bhubaneswar,
Argul, Jatni, Khurda, Odisha 752050, India

 $^{c}$  {Department of Physics, Faculty of
Science, University of Hradec Kr\'{a}lov\'{e}, Rokitansk\'{e}ho 62,
50003 Hradec Kr\'{a}lov\'{e},
 Czech Republic}

 $^{d}$
{The Czech Academy of Sciences,
 Nuclear Physics Institute,
 Hlavn\'{\i} 130,
250 68 \v{R}e\v{z}, Czech Republic}

$^{e}$
\textcolor{black}{ Institute of System Science, Durban University of Technology,
{4001}  Durban,
 South Africa}

 \newpage


\subsection*{Abstract}

The quantum-mechanical unitary-evolution
process of the loss of observability
at an exceptional-point singularity
is studied via an exactly-solvable example
tractable as a fermionic three-site generalization
of the popular bosonic Swanson model.


\subsection*{Keywords}

phenomenon of
non-Hermitian quantum degeneracies;

 \noindent
loss of observability via a
unitary-evolution process;

 \noindent
spike-shaped
corridors of
access to the singularity;

 \noindent
illustrative fermionic example;

\newpage

\section{Introduction}

In the paper
\cite{Styer}
which offers one of the most compact reviews of
the birth of quantum mechanics
we can read that
all of the available
``different formulations''
of the theory
are important because
in spite of being formally equivalent,
they
``provide different insights''
into multiple phenomena which ``seem weird \ldots
to our classical sensibilities''
\cite{Styer}.

One of these
reformulations of
the conventional quantum mechanics of unitary systems
became
known as ``quasi-Hermitian quantum mechanics''
(QHQM -- cf. its compact
introductory outline in section \ref{compa} below,
or some its more detailed reviews in \cite{Geyer,ali,book}).
It
provides, in particular, an entirely new insight in
various quantum analogues of the
classical
degeneracies and bifurcations (see, e.g.,
a brief discussion of the latter
topic in \cite{catast} or \cite{passage}).
The latter observation motivated also our present paper.
We will analyze and describe
the mechanism of quantum degeneracy in the form
provided by an elementary
though still nontrivial example.

First, we will
introduce our toy model in section \ref{themodel}.
The proof of existence
as well as a constructive
localization of the instant of
its quantum
degeneracy at the so called Kato's exceptional-point
singularity (EP, \cite{Kato})
will be then delivered
in section \ref{exceptionalpoint}.
We will show that
our model is characterized by a
rather fortunate combination of
phenomenological nontriviality
with a perceivable methodical appeal of
non-numerical tractability
enabling us to endow it with the
nontrivial, \textcolor{black}{higher-order}
EP of order three, EP=EP3
\textcolor{black}{(cf., e.g., \cite{St,EvaM})}.

The climax of our present message will then come
in sections \ref{primavera} and
\ref{corridort},
devoted to the
questions of
the behavior of the system
near the instant of the loss of its observability,
i.e., in the vicinity
of the EP3 singularity.
We will show
that
several related dangers
(including, first of all, the threat of instability)
can be kept under control since
in
every vicinity of the
singularity
there exists
a  ``corridor
of stability'',
i.e., a non-empty
unitary-evolution-supporting
subdomain of admissible parameters
with an
explicitly specified
boundary.

This will be our main mathematical result.
In the language of physics it will imply that
with the parameters lying
inside the latter corridor {\it alias\,}
``channel of unitarity'',
the reality of the spectrum, i.e.,
the stability
of the closed quantum system in question
will be guaranteed.
The demonstration of the existence
as well as of a fine-tuned
nature of the domain of stability
can be perceived as opening an
experimentally realizable
collapse
via a unitary-evolution
fall of the quantum system in question
into its EP3 singularity.

A brief summary of these results will be added in section \ref{summary}.


\section{The concept of quasi-Hermiticity in quantum theory\label{compa}}

In Schr\"{o}dinger picture \cite{Messiah},
the evolution of a stable
quantum system (also known as the ``closed'' quantum system)
is usually studied as a unitary process controlled, say,
by a time-independent Hamiltonian $H$
and by the time-dependent Schr\"{o}dinger equation
for states $|\psi(t)\kt$ in Hilbert space ${\cal H}_{\rm (physical)}$,
 \be
 {\rm i}\partial_t|\psi(t)\kt = H\,|\psi(t)\kt\,.
 \label{tdse}
 \ee
In 1992, Scholtz, Geyer and Hahne \cite{Geyer}
pointed out that in the context of applications
one can either use
the conventional version
of the theory
(in which $H$ is assumed self-adjoint
in a suitable
preselected
Hilbert space ${\cal H}_{\rm (mathematical)}={\cal H}_{\rm (physical)}$),
or its equivalent ``quasi-Hermitian'' reformulation
in which one admits that $H\neq H^\dagger$
in ${\cal H}_{\rm (mathematical)}$ since
${\cal H}_{\rm (mathematical)}\neq {\cal H}_{\rm (physical)}$
(see more details, for example, in \cite{book}).

A comprehensive explanation of the
apparent paradox can be found in multiple
physics-oriented reviews
\cite{Geyer,ali,Carl}.
The essence of the idea is
in fact rather elementary:
Besides the ``conventional''
physical Hilbert space ${\cal H}_{\rm (physical)}$
one simply introduces
another, auxiliary, unitarily non-equivalent and
manifestly unphysical but still
persuasively more user-friendly
Hilbert space ${\cal H}_{\rm (mathematical)}$.
The key point is that
all of the representations and calculations
are only being performed
in the latter space.
This means that in this space one encounters
the non-Hermiticity $H\neq H^\dagger$,
knowing that it
does not play any role at all.
The
probabilistic interpretation
and predictions are only obtainable via the so-called Dyson-map
correspondence \cite{Dyson} between ${\cal H}_{\rm (mathematical)}$
(in which the quantum system in question is represented
mathematically)
and
${\cal H}_{\rm (physical)}$
(only in which the questions of physics are treated correctly).

One of the main technical grounds of the appeal of the
``two-Hilbert-space'' reformulation of quantum mechanics
(in Schr\"{o}dinger picture)
is the fact that the (in practice, prohibitively
complicated) ``physical''  unitarity of the
evolution (in ${\cal H}_{\rm (physical)}$)
is {\em equivalently\,} re-expressed as a
so-called quasi-unitarity
of the evolution in ${\cal H}_{\rm (mathematical)}$.
In other words, the Stone-theorem-related
(but, by assumption, technically prohibitively
complicated)
Hermiticity of $H$
in ${\cal H}_{\rm (physical)}$
is simply given,
in the preferred
and friendlier ``mathematical representation'' Hilbert space
${\cal H}_{\rm (mathematical)}$, the next-to-elementary form
of constraint
 \be
 H^\dagger \Theta=\Theta\,H
 \textcolor{black}{\,,\ \ \ \
 \Theta=\Theta^\dagger > 0 }
 \label{quaquas}
 \ee
imposed in the working Hilbert space
${\cal H}={\cal H}_{\rm (mathematical)}$
and called, by mathematicians \cite{book,Dieudonne},
the quasi-Hermiticity of $H$.


\section{Swanson-like finite-matrix models\label{themodel}}

In the context of quantum phenomenology,
our present basic inspiration comes from
Swanson's study \cite{Swanson}
of the \textcolor{black}{manifestly non-Hermitian}
three-parametric bosonic
toy-model Hamiltonian
 \be
 H= \omega\, a^\dagger a
    + \alpha \,a^2 + \beta\, (a^\dagger )^2
    \textcolor{black}{\,,\ \ \ \
\omega\,,\alpha\,,\beta \in \mathbb{R}\,
    ,\ \ \ \
    \alpha\neq \beta }
 \ee
(with annihilation operator $a$
and creation operator $a^\dagger$)
and from its extension
to a fermionic two-site scenario as described in paper \cite{Bijan}.

\subsection{Elementary fermionic Hamiltonian in a two-site arrangement\label{tootsie}}

The most elementary fermionic Swanson-like
quantum Hamiltonian of paper \cite{Bijan} has the form
\begin{equation}
\label{eq:HF-fermionic}
{H} = \omega\, c_1^\dagger c_1 + (1-\omega)\,c_2^\dagger c_2
    + \beta\, c_1^\dagger c_2^\dagger + \alpha \,c_2 c_1\,,\ \ \ \
\omega\,,\alpha\,,\beta \in \mathbb{R},
\end{equation}
\textcolor{black}{can be kept Hermitian iff $\alpha=\beta$,}
and is defined \textcolor{black}{again} in terms
of the two pairs of the
creation and annihilation operators
satisfying the standard
anticommutation relations
\begin{equation}
\{c_j,c_k^\dagger\} = \delta_{jk},
\qquad
\{c_j,c_k\} = 0 = \{c_j^\dagger,c_k^\dagger\}
\qquad j,k = 1,2\,.
\end{equation}
The corresponding full-fledged fermionic Hilbert space
(or rather the Fock space \textcolor{black}{denoted,
say, by symbol ${\cal H}^{(\infty)}$})
was projected, in paper \cite{Bijan},
to its four-dimensional subspace ${\cal H}^{(4)}$
spanned by the vacuum $|0\rangle $
(such that
$
c_1|0\rangle = 0
$ and $
c_2|0\rangle = 0
$)
and by the other three basis states
$
|1\rangle = c_1^\dagger\,|0\rangle$,
$|{2}\rangle = c_2^\dagger\,|0\rangle$ and
$|3\rangle = c_1^\dagger c_2^\dagger\,|0\rangle\,$.%
%
%
%
%
%
%
%

The truncation of the space implied
a simplification $H \to \widetilde{H}^{(4)}$
such that
\begin{equation}
\widetilde{H}^{(4)}|0\rangle = \beta |3\rangle\,,\ \ \
\widetilde{H}^{(4)}|1\rangle = \omega |1\rangle\,,\ \ \
\widetilde{H}^{(4)}|2\rangle = (1-\omega) |2\rangle\,,\ \ \
\widetilde{H}^{(4)}|3\rangle = \alpha |0\rangle + |3\rangle\,.
\end{equation}
Then, it was possible to rewrite
the
truncated (i.e.,
tilded) Hamiltonian in the
partitioned finite-matrix form
tractable as a direct sum
of its two independent two-by-two-matrix
components,
\begin{equation}
\label{eq:H-matrix-final}
\widetilde{H}^{(4)}=
\left (
\begin{array}{c|cc|c}
 0 & 0 & 0 & \alpha  \\
 \hline
 0 & \omega  & 0 & 0 \\
 0 & 0 & (1-\omega)  & 0 \\
 \hline
 \beta  & 0 & 0 & 1 \\
\end{array}
\right )
=\left (
\begin{array}{cc}
\omega  & 0  \\
 0 & (1-\omega)  \\
\end{array}
\right )
\bigoplus
\widetilde{H}^{(2)}(\alpha,\beta)\,.
\end{equation}
One of the components was trivial (i.e., Hermitian and diagonal)
so that the authors of paper \cite{Bijan}
could restrict their attention to the
real and manifestly
non-Hermitian submatrix
\be
\widetilde{H}^{(2)}(\alpha,\beta)=\left (
\begin{array}{cc}
0& \alpha  \\
 \beta & 1   \\
\end{array}
\right )\,,
\ \ \ \ \ \alpha \neq \beta
\,.
\label{ma2}
\ee
This was an important simplification
\textcolor{black}{(notice that
via the attached inequality
these authors have already
excluded the
manifestly Hermitian special case as
not sufficiently interesting)} which
facilitated the
search for a non-Hermitian
exceptional-point degeneracy \cite{Berry,Heiss,Heissb}.

The latter analysis served also as an inspiration
and methodical guide of our present paper.
The essence of the approach
can be summarized as follows.
Firstly, matrix (\ref{ma2}) is to be assigned its
two bound-state-energy-mimicking eigenvalues
 \be
 E_\pm^{(2)} (\alpha,\beta)=\frac{1}{2}\,
 \left [1 \pm \sqrt{1+4\alpha\beta}
 \right ]\,.
 \label{forum}
 \ee
This implies that
the spectrum will cease to be real,
i.e., that
the
evolution of the quantum system in question will cease to be
unitary -- i.e., it will cease
to be of our present interest -- whenever  $\alpha\beta< -1/4$.

\subsection{Exceptional-point loss of observability \textcolor{black}{in
the two-site arrangement}\label{ootsie}}

At the first sight
it is less obvious that the boundary $\alpha\beta=-1/4$
of the latter unphysical interval
remains unphysical.
In
the limit $\alpha\beta\to -1/4$,
indeed, matrix (\ref{ma2}) ceases to be diagonalizable
and, as a consequence, it possesses just a single eigenvector.
In the language of linear algebra (cf., e.g., chapter 2
of Kato's monograph \cite{Kato}),
such a limit can be called an exceptional point (EP)
and, in this special case,
an exceptional point of order two, EP=EP2.

For the Hermitian
$N$ by $N$ matrices $M^{(N)}$
the EP singularities
cannot exist at all.
Indeed, any such a matrix
can be
diagonalized
using a suitable unitary matrix (say, ${\cal U}$).
The latter transformation matrices
can be then formally
perceived as composed of the columns formed by
the eigenvectors of $M^{(N)}$.

This is the idea which can partially be transferred to
non-Hermitian matrices $H^{(N)}$
and even to their singular EP limits.
In the latter case, the
canonical result of such a
transformation is widely accepted to be the
block-diagonal matrix containing one (or, perhaps, several)
bidiagonal $K$ by $K$ Jordan-block submatrices
 $$
 J^{(K)}(E_0)=
 \left[ \begin {array}{ccccc}
                     E_0&1&0&\ldots&0
 \\\noalign{\medskip}0&E_0&1&\ddots&\vdots
 \\\noalign{\medskip}0&0&E_0&\ddots&0
 \\\noalign{\medskip}\vdots&\ddots&\ddots&\ddots &1
 \\\noalign{\medskip}0&\ldots&0&0&E_0
 \end {array} \right]\,
 $$
in which the variable $E_0$ denotes the related
$K-$times degenerate
(i.e., EPK) eigenvalue.

In our specific two-by-two model (\ref{ma2})
let us abbreviate $\alpha^{(EP2)}=z$.
This means that
the one-parametric pencil
of our EP2-singular matrices acquires the one-parametric form
 $$
 \widetilde{H}^{(2)}(\alpha^{(EP2)}(z),\beta^{(EP2)}(z))=
 \left[ \begin {array}{cc} 0&z\\\noalign{\medskip}
 -1/(4\,{z}^{})&1\end {array} \right]
 $$
which can be assigned the
Jordan-block canonical representation $J^{(2)}(E_0)$ with $E_0=1/2$
(cf. Eq.~(\ref{forum})).
The construction can proceed
via the following, Schr\"{o}dinger-equation-resembling
relation
 \be
\widetilde{H}^{(2)}(\alpha^{(EP2)}(z),\beta^{(EP2)}(z))
\,
{\cal U}^{(EP2)}(z)
={\cal U}^{(EP2)}(z)
\,
J^{(2)}(1/2)\,.
 \label{byeq}
 \ee
The solution
 \be
 {\cal U}^{(EP2)}(z)=
 \left[ \begin {array}{cc} -1/2&1
\\\noalign{\medskip}-1/(4\,{z}^{})&0\end {array} \right]
 \ee
of this equation is a non-unitary
transformation matrix
called, usually, the transition matrix.

The latter formal analogue of ${\cal U}$
is still
defined as composed of certain
column vectors called associated vectors.
This has been thoroughly described in the paper \cite{Bijan}
which
will be used here as a
methodical guide.

For such a purpose
we will simplify
some of the formulae, mainly
via a reparametrization. This will
include an overall rescaling
of
matrix (\ref{ma2})
(by its multiplication by factor 2)
and
a constant shift of the energies.
\textcolor{black}{In this way, our} toy-model Hamiltonian
\textcolor{black}{$
\widetilde{H}^{(2)}(\alpha,\beta)$
of Eq.~(\ref{ma2})
becomes replaced by an}
untilded
traceless
Hamiltonian
 \be
 H^{(2)}=\left[ \begin {array}{cc} -1&{ {({A}^{2}-1)/z}}
\\\noalign{\medskip}z&1\end {array} \right]\,
\label{unti}
 \ee
\textcolor{black}{in which the readers can easily
check that $z=2\beta$ while $A^2=1+4\alpha\beta$.
At $A \neq 0$
and at all $z\neq 0$, both of the
shifted and non-degenerate}
bound-state energies remain
$z-$independent and equal to $\pm A$
so that for the sake of definiteness we can demand $A > 0$.

The \textcolor{black}{upgraded Hamiltonian (\ref{unti})
will still} have the two
independent
and arbitrarily normalized
eigenvectors forming, say, a two-by-two matrix
 \be
 \left[ \begin {array}{cc} 1-1/A&1+1/A
 \\\noalign{\medskip}z/A&-z/A\end {array} \right]\,.
 \ee
In the limit of $A\to 0$
we now reveal
the existence of the
EP2 singularity since the two columns of
the latter matrix cease to be linearly independent
in this limit.

In an alternative proof
of the EP2 nature of the $A \to 0$ degeneracy
we might directly recall
Eq.~(\ref{byeq})
and construct
its solution, say,
 \be
 {\cal U}^{(EP2)}(z)=\left[ \begin {array}{cc} -1&1
 \\\noalign{\medskip}z&0\end {array} \right]\,.
 \ee
The unconstrained real parameter $z$
enters such a transition matrix in an elementary manner.
In the context of
physics the variable $z$ is still a
relevant parameter since its value
can determine the
measurable characteristics of the quantum
system in question.

\textcolor{black}{In the context of physics the}
role of $z$ becomes particularly important
after one decides to study
the influence of perturbations -- cf. \cite{pertEP}.
In the \textcolor{black}{spirit of the considerations}
of our preceding section \ref{compa}
we may conclude
that in spite of its non-Hermiticity, our toy-model Hamiltonian
(\ref{eq:H-matrix-final}) admits the conventional physical
interpretation (i.e., a Hermitian or a quasi-Hermitian
stable-bound-states-generating status) if and only if $\alpha\beta> -1/4$.

\subsection{Swanson-like system in a more general three-site arrangement}

A straightforward extension
of the preceding considerations
to the three-site Hamiltonian
\begin{equation}
\label{eq:3HF-fermionic}
{H} = \sum_{j=1}^3\,\omega_j\, c_j^\dagger c_j +
    + \beta\, (c_1^\dagger c_2^\dagger +
    c_2^\dagger c_3^\dagger
     ) + \alpha\,     (c_2 c_1+
    c_3 c_2^{}
     )\,,\ \ \ \
\omega_j\,,\alpha\,,\beta \in \mathbb{R}\,,\ \
\alpha \neq \beta
\end{equation}
leads, {\it mutatis mutandis},
to an immediate three-term generalization
\begin{equation}
\label{eq:inal}
\widetilde{H}^{(8)}=\left (
\begin{array}{cc}
\omega_2  & 0  \\
 0 & \omega_1+\omega_3  \\
\end{array}
\right )
\bigoplus
\widetilde{H}^{(3)}_{[1]}(\alpha,\beta,\vec{\omega})
\bigoplus
\widetilde{H}^{(3)}_{[2]}(\alpha,\beta,\vec{\omega})\,
\end{equation}
of the two-site-related decomposition~(\ref{eq:H-matrix-final}).
\textcolor{black}{We are again omitting the
Hermitian-matrix limit with $\alpha=\beta$.
Hence, our assumption}
$\alpha \neq \beta$
\textcolor{black}{makes also
the two three-by-three}
real-matrix components
\be
\widetilde{H}^{(3)}_{[1]}(\alpha,\beta,\vec{\omega})=\left (
\begin{array}{ccc}
0& \beta & \beta \\
 \alpha &  \omega_1+\omega_2&0   \\
 \alpha & 0 &\omega_2+\omega_3   \\
\end{array}
\right )
\,\,
\label{ma2a}
\ee
and
\be
\widetilde{H}^{(3)}_{[2]}(\alpha,\beta,\vec{\omega})=\left (
\begin{array}{ccc}
\omega_1 & 0 & \beta \\
 0 &  \omega_3&\beta  \\
 \alpha & \alpha &\omega_1+\omega_2+\omega_3   \\
\end{array}
\right )
\,
\label{ma2b}
\ee
\textcolor{black}{of the Hamiltonian}
non-Hermitian.
After the different shifts $-\omega_2$
and  $-\omega_1-\omega_3$ of the
\textcolor{black}{respective subsystem's}
energy scales we get
\be
\widetilde{H}^{(3)}_{[1]}(\alpha,\beta,\vec{\omega})=\left (
\begin{array}{ccc}
-\omega_2& \beta & \beta \\
 \alpha &  \omega_1&0   \\
 \alpha & 0 &\omega_3   \\
\end{array}
\right )
\,\,
\label{ma2c}
\ee
and
\be
\widetilde{H}^{(3)}_{[2]}(\alpha,\beta,\vec{\omega})=\left (
\begin{array}{ccc}
-\omega_3 & 0 & \beta \\
 0 & - \omega_1&\beta  \\
 \alpha & \alpha &\omega_2   \\
\end{array}
\right )
\,.
\label{ma2d}
\ee
\textcolor{black}{We may conclude that both of these matrices are strictly
analogous to their two-by-two-matrix predecessor of Eq.~(\ref{ma2}).
In
the next step of our present considerations, in
other words, it will be necessary to
extend the
two-site
results of subsection \ref{ootsie}
to a suitable three-site generalization.}

\section{Exceptional-point degeneracy in the
three-site case\label{exceptionalpoint}}

Our study of the three-site toy-model
Hamiltonian $\widetilde{H}^{(3)}_{[1]}(\alpha,\beta,\vec{\omega})$
or
$\widetilde{H}^{(3)}_{[2]}(\alpha,\beta,\vec{\omega})$
may parallel the above-outlined two-site methodical guide.
After a suitable reordering of the basis one finds that
both of
the latter two matrices
\textcolor{black}{
(\ref{ma2c}) and (\ref{ma2d})}
may be given the same form.
Formally,
this will
enable us to analyze
just one of them, i.e., say,
matrix $\widetilde{H}^{(3)}_{[1]}(\alpha,\beta,\vec{\omega})$
of Eq. (\ref{ma2c}).

\subsection{An enrichment of the set of couplings}

Naturally, the parallels
\textcolor{black}{
between the two- and three-site cases}
are not strict.
In particular, for our present purpose of
enforcing a {\em complete\,} (i.e., EP3) collapse and
degeneracy
of the three-site system in question,
the number of the variable parameters available
in the
matrix of Eq.~(\ref{ma2c}) (or,
\textcolor{black}{if you wished, in the other
matrix of Eq.~}(\ref{ma2d}))
would be
insufficient.

A remedy is obvious. In the original Hamiltonian
of Eq.~(\ref{eq:3HF-fermionic})
one merely has to
enhance the flexibility of
the interactions and
\textcolor{black}{
to split the interaction terms by replacing
their original one-parametric version of Eq.~(\ref{eq:3HF-fermionic}) by a
two-parametric amendment yielding}
 \be
 \beta\, (c_1^\dagger c_2^\dagger +
    c_2^\dagger c_3^\dagger
     )\
     \to
     \
 \beta\, c_1^\dagger c_2^\dagger +\beta'\,
    c_2^\dagger c_3^\dagger
          \ee
and
     \be
      \alpha\,     (c_2 c_1+
    c_3 c_2^{}
     )\
     \to\
      \alpha\,     c_2 c_1+\alpha'\,
    c_3 c_2^{}\,.
 \ee
In
the resulting
amended form of decomposition~(\ref{eq:inal})
of the complete but reducible Hamiltonian,
one will still have to deal just with the
five-parametric
and, say, traceless real non-Hermitian
or quasi-Hermitian
submatrices.

\textcolor{black}{
This means that
we will have to use again
the same sequence of tricks as
described in detail above.
Thus, firstly, we will
have to make our
three-by-three matrix
traceless. This will be achieved
via a mere suitable shift of the
origin on the energy energy scale.
Secondly,
we will pre-multiply (i.e., rescale) the whole
amended matrix
$\widetilde{H^{(3)}}$
by such a constant that
its diagonal matrix elements
acquire the ``canonical'' form of
the triplet $\{-2,1-u,1+u\}$
containing a single variable $u$
(note that
no such auxiliary parameter was needed in the
two-site case
with the diagonal doublet $\{-1,1\}$
in Eq.~(\ref{unti})).}

\textcolor{black}{
The latter operation will
already lead to the mere trivial rescaling of the
off-diagonal matrix elements so that
the general untilded three-by-three-matrix
analogue of
its two-parameetric predecessor of Eq.~(\ref{unti})
will finally acquire the
following five-parametric form}
 \be
H=
  \left[ \begin {array}{ccc} -2&{ b}&d
\\\noalign{\medskip}a&1-u&0
\\\noalign{\medskip}{ c}&0&1+u\end {array} \right]\,.
 \label{mopmo}
 \ee
During a preliminary analysis
of the system we revealed that the
energy spectrum of our amended toy model (\ref{mopmo})
only depends on the products $ab=A$ and $cd=B$
so that we just reparametrized the matrix
\textcolor{black}{
and obtained its final form},
 \be
  H(a,A,B,d,u)=
\left[ \begin {array}{ccc} -2&{ {A}/{a}}&d
\\\noalign{\medskip}a&1-u&0
\\\noalign{\medskip}{ {B}/{d}}&0&1+u\end {array} \right]\,.
 \label{mobmo}
 \ee
\textcolor{black}{
Needless to add that the two initial choices of the
tilded matrix  $\widetilde{H}^{(3)}_{[1]}(\alpha,\beta,\vec{\omega})$
or
$\widetilde{H}^{(3)}_{[2]}(\alpha,\beta,\vec{\omega})$
lead to the two different (but, in both cases,
entirely straightforward) forms of
the redefinitions $\{\alpha,\alpha',\beta,\beta',
\omega_1,\omega_2,\omega_3\} \to \{a,d,A,B,u\}$
of the parameters.
In both of these cases, in any case,
it makes sense to work with the more
compact new set of parameters for which
it is easy to deduce the}
related secular equation
 $$
{{\it E}}^{3}- \left( B+3+{u}^{2}+A \right) {\it E}+B-Bu+A+2+uA-2\,{
u}^{2}=0\,.
 $$
This equation and, hence, all of its three (real or complex)
energy roots are merely three-parametric
and $a-$ and $d-$independent.

\subsection{The instant of degeneracy}

The latter feature
of the energies $E_j$
representing either a stable bound state or an
unstable resonance
at every particular
subscript $j=1,2$ or $3$
is a decisive technical advantage of the model.
For a guarantee of a
specific triple degeneracy $E_j \to E^{(EP3)}=0$
of our present interest
it is, in the light of the secular equation,
necessary -- though not necessarily
sufficient --  to satisfy the following two constraints,
 $$
 B+3+{u}^{2}+A =0\,,\ \ \ \
 B-Bu+A+2+uA-2\,{u}^{2}=0\,,
 $$
i.e., a pair of
the coupled algebraic polynomial equations.

\begin{figure}[h]                    
\begin{center}                         
\epsfig{file=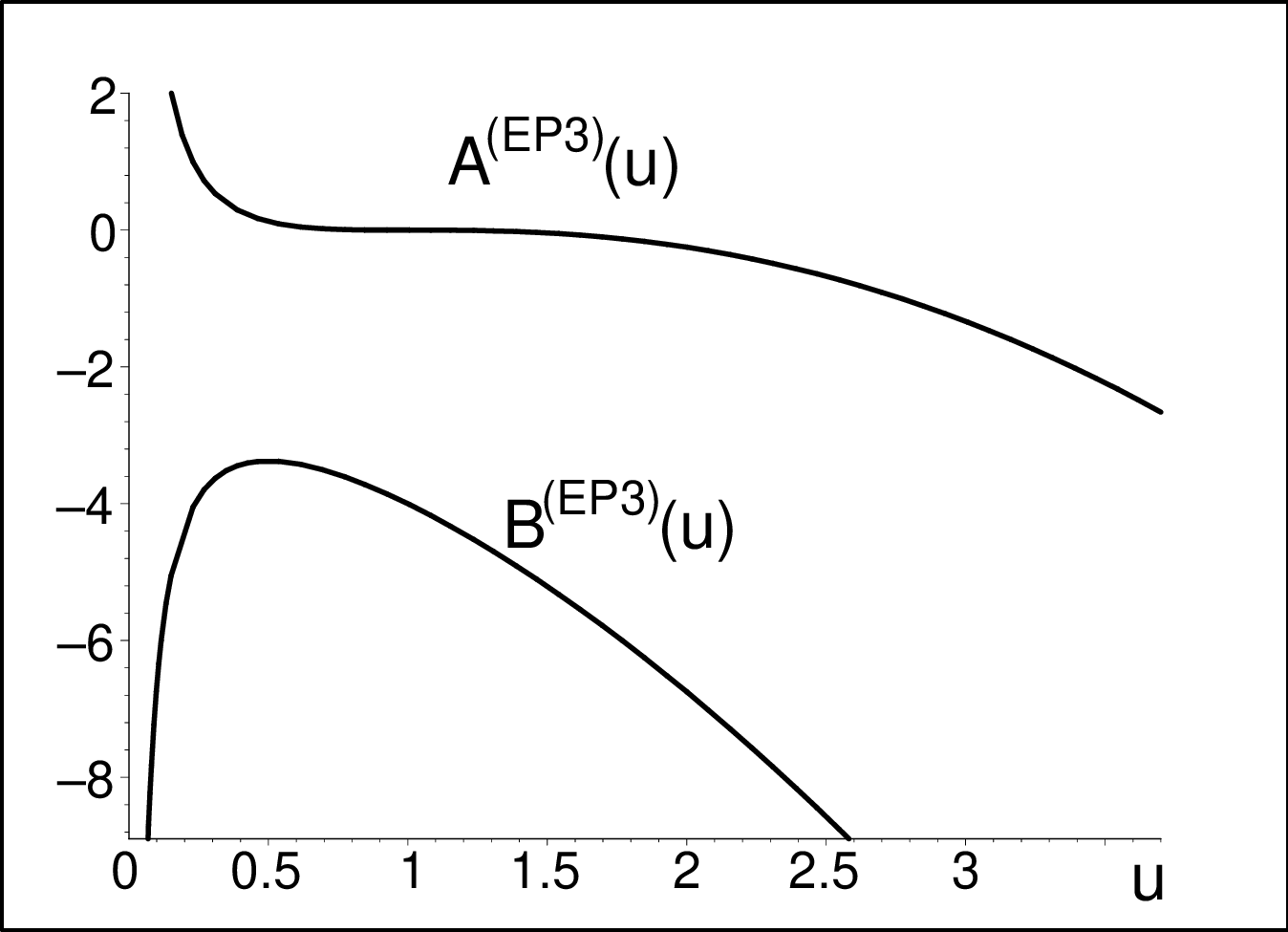,angle=270,width=0.35\textwidth}
\end{center}    
\caption{The pair of functions $A^{(EP3)}(u)$ and $B^{(EP3)}(u)$ at $u>0$.
 \label{Qws5}}
\end{figure}

During their analysis one reveals that for any predetermined and,
presumably, variable
``physical'' parameter $A$ (or $B$),
the corresponding
EP3-localization functions $u^{(EP3)}(A)$
and $B^{(EP3)}(A)$
(or, alternatively, $u^{(EP3)}(B)$
and $A^{(EP3)}(B)$)
will only be available
in a closed but highly impractical form of Cardano formula.
Fortunately, both of the EP3-determining
equations are just linear in $A$ and $B$.
For this reason, it makes sense to
treat $u$ as  a variable parameter determining
the following closed-form
EP3 singularity localization with
 \be
 A=A^{(EP3)}(u)=
  -\frac{1}{2}\,
  (u^2-3u+3)+\frac {1}{2u}
 =
 \frac{(1-u)^3}{2u}
 \,
 \label{mocmo}
 \ee
and
 \be
 B=B^{(EP3)}(u)=
  -\frac{1}{2}\,
  (u^2+3u+3)-\frac {1}{2u}
 =
 -\frac{(1+u)^3}{2u}=A^{(EP3)}(-u)\,.
 \label{modmo}
 \ee
These formulae remain elementary and user-friendly.
Both of these curves are
smooth and do not intersect.
They are also both asymptotically deceasing
to minus infinity
while,
near the origin, their growth $\sim u^{-1}$ (i.e., to infinity
again)
proceeds in opposite directions (see Figure \ref{Qws5}).

\subsection{Parallelization of eigenvectors}

An explicit demonstration that the energy degeneracy
is really Kato's EP3
singularity
requires the proof
that
the
phenomenon of
degeneracy involves
not only the spectrum but
also the eigenvectors.
For such a proof we have to
choose and fix an arbitrary suitable value of
the free parameter $u=u^{(EP3)}$, and we then have to
evaluate the related, EP3-specifying functions
(\ref{mocmo}) and
(\ref{modmo}).
For illustration,
let us pick up, say,
 \be
 u^{(EP3)}_{(sample)}=5\,.
 \label{pickup}
 \ee
This will enable us to
deduce that
 $$
 A^{(EP3)}_{(sample)}=-32/5\,,
 \ \ \ \ \
 B^{(EP3)}_{(sample)}= -108/5\,.
 $$
These values are negative
and not too small
in a way compatible with Figure \ref{Qws5}.
\textcolor{black}{The
elementary (i.e., just one-parametric)
nature of the two curves as depicted in
Figure \ref{Qws5}
also enables us to see that
the EP3 singularity does exist
for an arbitrary choice of the positive
variable parameter $u$.}

Now we may recall the
\textcolor{black}{
next step of the}
methodical guide
as provided
by subsection \ref{tootsie},
and we may construct
a suitable 
EP3-related transition matrix ${\cal U}^{(EP3)}$.
This matrix should be treated as an analogue of its
predecessor ${\cal U}^{(EP2)}(z)$
as defined by Eq.~(\ref{byeq}) above.
We just have to replace the latter equation by its
EP3-related three-site descendant
 \be
{H}^{(3)}(a,A^{(EP3)}(u^{(EP3)}),B^{(EP3)}(u^{(EP3)}),d,u^{(EP3)})
\,
{\cal U}^{(EP3)}
={\cal U}^{(EP3)}
\,
J^{(3)}(0)\,.
 \label{byeq3}
 \ee
The net result of the application of this approach
in the scenario of Eq.~(\ref{pickup}) with
$u^{(EP3)}_{(sample)}=5$
is immediate: We
find
that the canonical form
 \be
 J^{(3)}(0)=\left[ \begin {array}{ccc} 0&1&0\\\noalign{\medskip}0&0&1
 \\\noalign{\medskip}0&0&0\end {array} \right]
 \label{tenham}
 \ee
of our singular EP3 Hamiltonian is obtained using the
transition matrix which is, incidentally, $a-$ and $d-$dependent,
 \be
 {\cal U}^{(EP3)}=
 \left[ \begin {array}{ccc} -24&-2&1\\\noalign{\medskip}-6\,a&a&0
 \\\noalign{\medskip}-{ {432}}/({5}{d})
 &-{ {108}}/({5}{d})&0\end {array} \right]\,.
 \label{eq3}
 \ee
The existence of such a solution of Eq.~(\ref{byeq3}) implies that
the proof of the EP3 status of the degeneracy is completed.

\section{Collapses to the EP3 degeneracy and/or its unfoldings\label{primavera}}

Our main concern is the
study of
non-Hermitian but quasi-Hermitian
Hamiltonian matrices
$H(a,A,B,d,u)$
of Eq.~(\ref{mobmo})
in which the values of the
lower-case parameters remain unconstrained,
and in which the spectrum happens to be independent of $a$ and $b$.

Up to now we revealed that in the singular EP3 limit
with
$A=A^{(EP3)}(u)$
(cf. Eq.~(\ref{mocmo}))
and with
$B=B^{(EP3)}(u)$
(cf. Eq.~(\ref{modmo}))
it makes sense to remove the
surviving
$u-$dependence ambiguity by
picking up any toy-model
value of $u=u^{(EP3)}$.
At this value,
the spectrum becomes degenerate
while
the Hamiltonian itself ceases to be diagonalizable,
i.e., observable and phenomenologically meaningful.
This means that our model may re-acquire
its conventional physical meaning and
probabilistic interpretation only after
we
change at least one of its three independent EP3 parameters.

\subsection{Generic case: the emergence of resonances near the EP3 extreme\label{sub}}

Once we construct, at any preselected $u=u^{(EP3)}$, the
transition-matrix
solution ${\cal U}^{(EP3)}$
of Eq.~(\ref{byeq3})
(which is to be assumed invertible),
we can re-interpret such a matrix as
a means of transformation of the Hamiltonian,
 \be
{H}^{(3)}(a,A^{(EP3)}(u^{(EP3)}),B^{(EP3)}(u^{(EP3)}),d,u^{(EP3)})
={\cal U}^{(EP3)}
\,\mathfrak{h}^{(3)}{(0)}\,
\left [ {\cal U}^{(EP3)}
\right ]^{-1}\,.
 \label{byeq3b}
 \ee
Here we changed the denotation of
$J^{(3)}(0)=\mathfrak{h}^{(3)}{(0)}$
in order to
emphasize that such a Jordan matrix
has to be
reinterpreted
as a modified, isospectral version of the Hamiltonian.

In the next step the symbol $\mathfrak{h}^{(3)}{(0)}$
is to be further reinterpreted as an unperturbed
limit of its general perturbed extension
 \be
 \mathfrak{h}^{(3)}{(\lambda)}=
 \left[ \begin {array}{ccc} 0&1&0\\\noalign{\medskip}0&0&1
 \\\noalign{\medskip}0&0&0\end {array} \right]
 +
 \lambda\,
 \left[ \begin {array}{ccc} V_{1,1}&V_{1,2}&V_{1,3}
 \\\noalign{\medskip}V_{2,1}&V_{2,2}&V_{2,3}
 \\\noalign{\medskip}V_{3,1}&V_{3,2}&V_{3,3}
 \end {array} \right]\,.
 \label{m[30]}
 \ee
The auxiliary variable $\lambda \neq 0$ is to be assumed,
in the spirit of the conventional perturbation-theory
considerations \cite{Kato},
sufficiently small.
Interested readers can
find the method of construction as well as a subsequent
discussion of the solution of the related
perturbed Schr\"{o}dinger equation
 \be
 \mathfrak{h}^{(3)}{(\lambda)}\, |\psi_j(\lambda) \kt
 = E_j(\lambda)\,|\psi_j(\lambda) \kt\,,
 \ \ \ j = 1,2,3
 \ee
in section III.C of the paper \cite{admissible}.

During the present stage of development it makes sense
to mention only
that the latter general study
came to
a rather unexpected and
seemingly paradoxical
unavoidable
conclusion
that
once we admit
perturbations $\lambda V$ possessing a
non-vanishing particular
matrix element
$V_{3,1}\neq 0$,
the resulting perturbed spectrum $\{E_j(\lambda)\}$
can never be real at small $\lambda$s.
From a purely pragmatic perspective
this means that in the generic, unconstrained
dynamical regime, at least some of
the states
near the EP3 singularity become unstable
(i.e., resonances).
This will happen
irrespectively of the size of
parameter~$\lambda$.

In more detail, we are going to return to this
observation, in its
explicit application to our present model,
in section \ref{corridort} below.
Preliminarily,
it is just sufficient to say now that
the abstract form of the puzzle
will find a clarification and resolution in
the necessity of
using a generalized,
parameter-dependent form of
the critical
matrix element $V_{3,1}= V_{3,1}(\lambda)$
which
must be
such
that $\lim_{\lambda \to 0}V_{3,1}(\lambda)= 0$.
This
is a rather unusual form of
constraint which has to be interpreted as a part of
a necessary innovation of perturbation
theory
in the quasi-Hermitian quantum-mechanics
framework \cite{pertsym}.
In the present specific-system context,
a detailed explanation of the paradox
is to be found below.

\subsection{Admissible perturbations and quasi-Hermitian unfoldings\label{subrb}}

At any convenient value of the EP3-characterizing parameter $u=u^{(EP3)}$
we may construct the transition matrix ${\cal U}^{(EP3)}$
by solving Eq.~(\ref{byeq3}).
This will enable us to
transform
the singular EP3 limit of our Hamiltonian
into its
canonical Jordan-matrix form
$\mathfrak{h}^{(3)}{(0)}=J^{(3)}(0)$
as given by Eq.~(\ref{tenham}).

In opposite direction, we will now
replace the EP3 Hamiltonian
$\mathfrak{h}^{(3)}{(0)}$
by its perturbed lower-case descendant
$\mathfrak{h}^{(3)}{(\lambda)}$
of Eq.~(\ref{m[30]}).
Simultaneously, we will make
use of our knowledge of the
same, $\lambda-$independent
transformation matrix ${\cal U}^{(EP3)}$
as a means of definition of another, new and isospectral
$\lambda-$dependent
Hamiltonian
 \be
 {H}^{(3)}(a(\lambda),A(\lambda),B(\lambda),
 d(\lambda),u(\lambda),\lambda)
 ={\cal U}^{(EP3)}
 \,
 \mathfrak{h}^{(3)}{(\lambda)}\,
 \left [{\cal U}^{(EP3)}\right ]^{-1}
 \,.
 \label{byeq3z}
 \ee
Due to its definition, the latter right-hand-side matrix
is
tractable, naturally, as a perturbed form
of its singular EP3 limit.
Incidentally, its symbol contains one more variable (viz., parameter
$\lambda$ itself) because the zero matrix elements of
its unperturbed predecessor
may now be expected to be replaced by some non-vanishing,
perturbation-proportional expressions of course.

Let us now recall that when
we
analyzed the
influence of perturbations
in paragraph \ref{sub},
we observed that in the generic case, not all of
the three perturbed, EP3-unfolding energies
remained real.
This means that
in the quasi-Hermitian, more restrictive dynamical regime
characterized
by the unitarity of the evolution as well as
by the reality
of the spectrum (i.e., by the stable
bound-state interpretation of the spectrum),
strictly speaking,
the process of the unfolding of the EP3 singularity
must be, somehow, fine-tuned.

A detailed form of such a key physics-
and stability-motivated guarantee of the
reality of all of the three branches of the perturbed energy
near the instant of its EP3 degeneracy
has already been studied
in paper \cite{corridors}.
For the three-by-three matrices,
in particular,
the main result of the analysis was
formulated there as Lemma Nr. 1.

\textcolor{black}{For our present purposes it is fully sufficient to
recall this result which, in essence, states that
the smallness of the perturbations near the EP3 singularity
should not be measured by the norm of the perturbation operator but
rather by
certain individual weights
of the separate matrix elements of the perturbation matrix.
Incidentally, interested readers may find
a very good introductory guide to the reasons
of such a mathematical subtlety
in paper \cite{EvaM}.}

\textcolor{black}{
For the needs of these readers
let us only add here that the Lemma in question,
in essence,}
stated that
in the vicinity of the EP3 singularity with $E^{(EP3)}=0$
one has to work
with a new {\it ad hoc\,} measure of smallness
$\varrho= |\lambda^{1/3}|$ in place of the parameter $\lambda$
itself.
Then, it was shown that
the perturbed energy spectrum
will consist of the three ${\cal O}(\varrho)$ branches
$E_j=\varrho\,\varepsilon_j$, $j=1,2,3$
which remain non-vanishing and real
if and
only if
we replace the naive
perturbed-Hamiltonian ansatz (\ref{m[30]})
by its ``re-gauged''
alternative, say, of
the most elementary three-parametric form
 \be
 \mathfrak{h}^{(3)}{(\varrho)}=
 \left[ \begin {array}{ccc} 0&1&0\\\noalign{\medskip}0&0&1
 \\\noalign{\medskip}0&0&0\end {array} \right]
 +
 \left[ \begin {array}{ccc} 0&0&0
 \\\noalign{\medskip}\varrho^2\,W_{2,1}&0&0
 \\\noalign{\medskip}\varrho^3\,W_{3,1}&\varrho^2\,W_{3,2}&0
 \end {array} \right]\,.
 \label{m[30]b}
 \ee
We may abbreviate
$W_{2,1}=\alpha$,
$W_{3,2}=\beta-\alpha$ and
$W_{3,1}=\gamma$
and arrive at
the standard secular equation which will acquire
the following, slightly unusual form
  $$
  \det\,\left[ \begin {array}{ccc} -\varrho\,\epsilon&1&0
  \\\noalign{\medskip}{\varrho}^{2}\,\alpha&-\varrho\,\epsilon&1
  \\\noalign{\medskip}{\varrho}^{3}
  \gamma&{\varrho}^{2}(\beta-\alpha)&-\varrho\,\epsilon
  \end {array} \right]=
  (\epsilon^3-\beta\,\epsilon-\gamma)\,\varrho^3=
  0\,.
  $$
This enables us to conclude that
the
unfolding of the spectrum
will proceed via three real branches
if and only if the
three
$\varrho-$independent
roots of equation
$\epsilon^3-\beta\,\epsilon-\gamma=0$
will be non-vanishing and real.

The related qualitative analysis of the latter equation
is straightforward. Firstly, we must have
 \be
 \beta>0\,.
 \label{subvic1}
 \ee
Otherwise, we would have just one real root.
Secondly, it is equally straightforward to deduce that the value of
the other parameter $\gamma$
must be localized inside the interval of
 \be
 \gamma \in (-2\,z^3,2\,z^3)\,,\ \ \ \ z=\sqrt{\beta/3}
 \label{subvic2}
 \ee
(see {\it loc. cit.} for a detailed proof).
Thus, we may summarize
that
the survival of the reality of the spectrum
during the unfolding
(or, equivalently, during the
process of the fall of the
system to its EP3 singularity)
can be perceived as
guaranteed by the
restriction of the variability of parameters
just to
the above-constructed
well-specified sub-vicinity
of their EP3 limit
as specified by Eqs.~(\ref{subvic1}) and (\ref{subvic2}).

\section{Discussion\label{corridort}}

During
some applications of
our innovative
and above-outlined perturbation-analysis approach
we revealed that after a transition to
numerical calculations,
the control of precision
must
be done with due care.
\textcolor{black}{
In this sense, our present
observations may be compared with the
results of multiple other recent studies,
both theoretical and experimental, in which
the relevance of the
occurrence of the
exceptional points
has been found to range from the physics
of ultracold atoms \cite{SSHb}
up to the
descriptions of the physical
systems with nonlinear dynamics \cite{Nimroda}.}

\textcolor{black}{In all of those contexts, the localization of the
EP singularities is challenging in a way well sampled also by our
present}
choice of $u^{(EP3)}_{(sample)}=5$ for which
we noticed that
some of the aspects of
the behavior of the system near its EP3 singularity
may in fact remain invisible, in particular,
due to an insufficient precision of the computer arithmetics.

Let us now add a few further notes on this topic.


\subsection{Closely avoided, false level crossings
near the EP3 degeneracy\label{subr}}

In support
of the latter words of warning, our
first comment will concern the apparent
simplicity of the model.
One might expect that
whenever one assumes that
$u \neq u^{(EP3)}_{}$,
the evaluation and qualitative description of the
spectrum
becomes routine.
Unfortunately, it is not always so.
For explanation,
let us
return to the choice of Eq.~(\ref{pickup}) with
 $
 u^{(EP3)}_{}=5
 $,
and
let us study
the sample version of
our model with $u \neq  u^{(EP3)}_{}$
and with the spectrum determined
by the polynomial
secular equation
 \be
 {{\it E}}^{3}- \left( -25+{u}^{2} \right) {\it E}-26+76
 \,u/5-2\,{u}^{2}=0\,.
 \label{toysek}
 \ee
Its closed-form solvability
in terms of Cardano formulae
would yield the
$u-$parametrized
spectrum in an arbitrary
(one-dimensional) vicinity
of the EP3 singularity.

In principle, these
solutions are algebraic and exact
but in practice,
their use is not too comfortable.
The reason is that
they are formally expressed
as sums of complex numbers. Thus,
due to the ubiquitous numerical errors,
even some of
the real energy values $E_j(u)$
may happen to be (wrongly)
reclassified as complex.

\begin{figure}[h]                    
\begin{center}                         
\epsfig{file=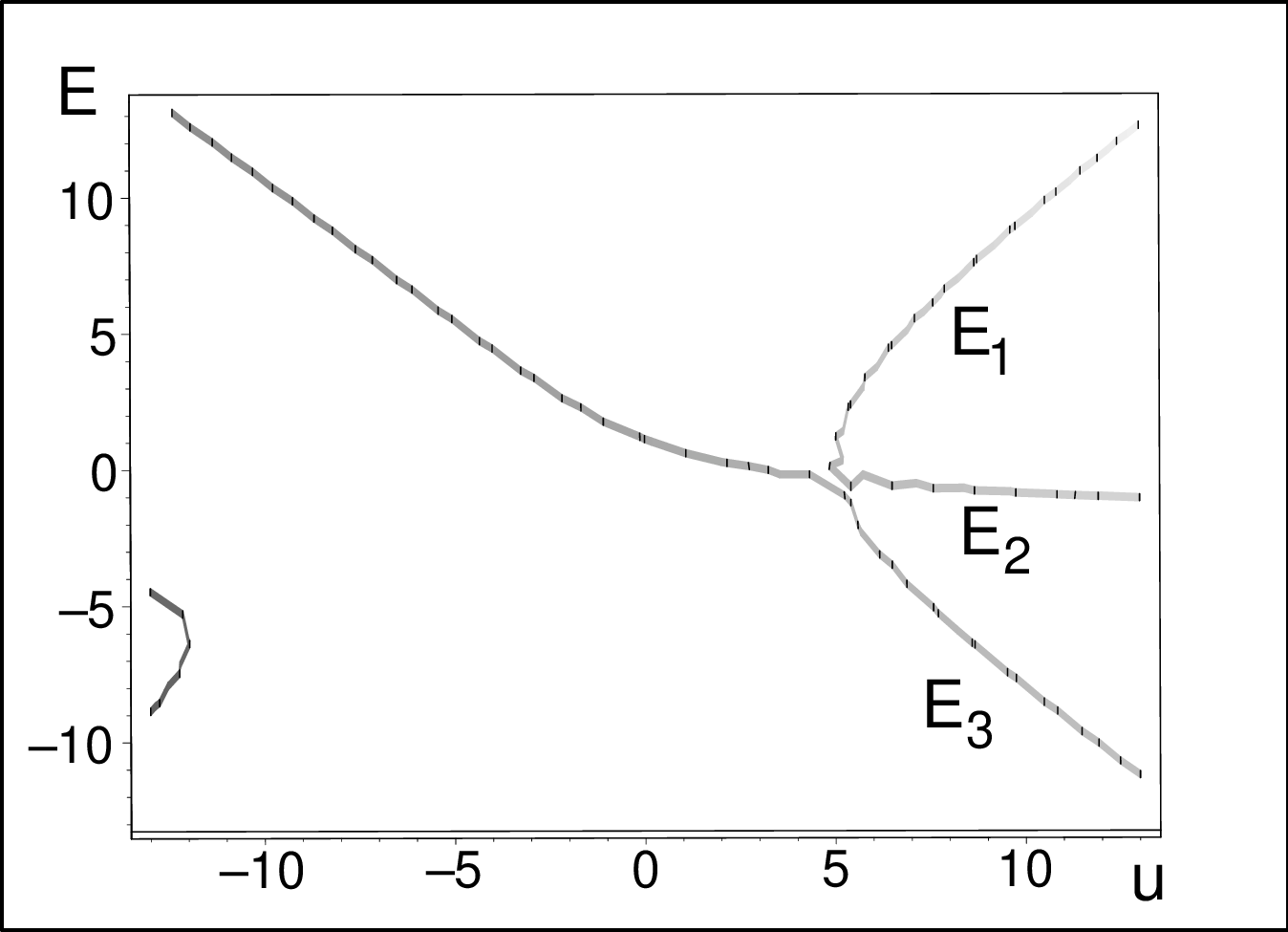,angle=270,width=0.35\textwidth}
\end{center}    
\caption{Numerical localization of the real (i.e., bound-state) energy levels
suggesting their degeneracy near $u \approx
u^{(EP3)}=5$.
 \label{Qwer5s}}
\end{figure}

Alternatively,
it is possible to use a more robust and
purely
numerical
localization of the
real eigenvalues, with a typical result
displayed
here in Figure \ref{Qwer5s}.
Intuitively, such an approximate graphical result seems
appealing and
fully compatible with
a tentative hypothesis that the EP3 singularity
(localized at $E=0$)
does exist, and
that it manifests itself as an unavoided crossing of
the triplet of levels $\{E_1(u),E_2(u),E_3(u)\}$ at $u=5$.

\begin{figure}[h]                    
\begin{center}                         
\epsfig{file=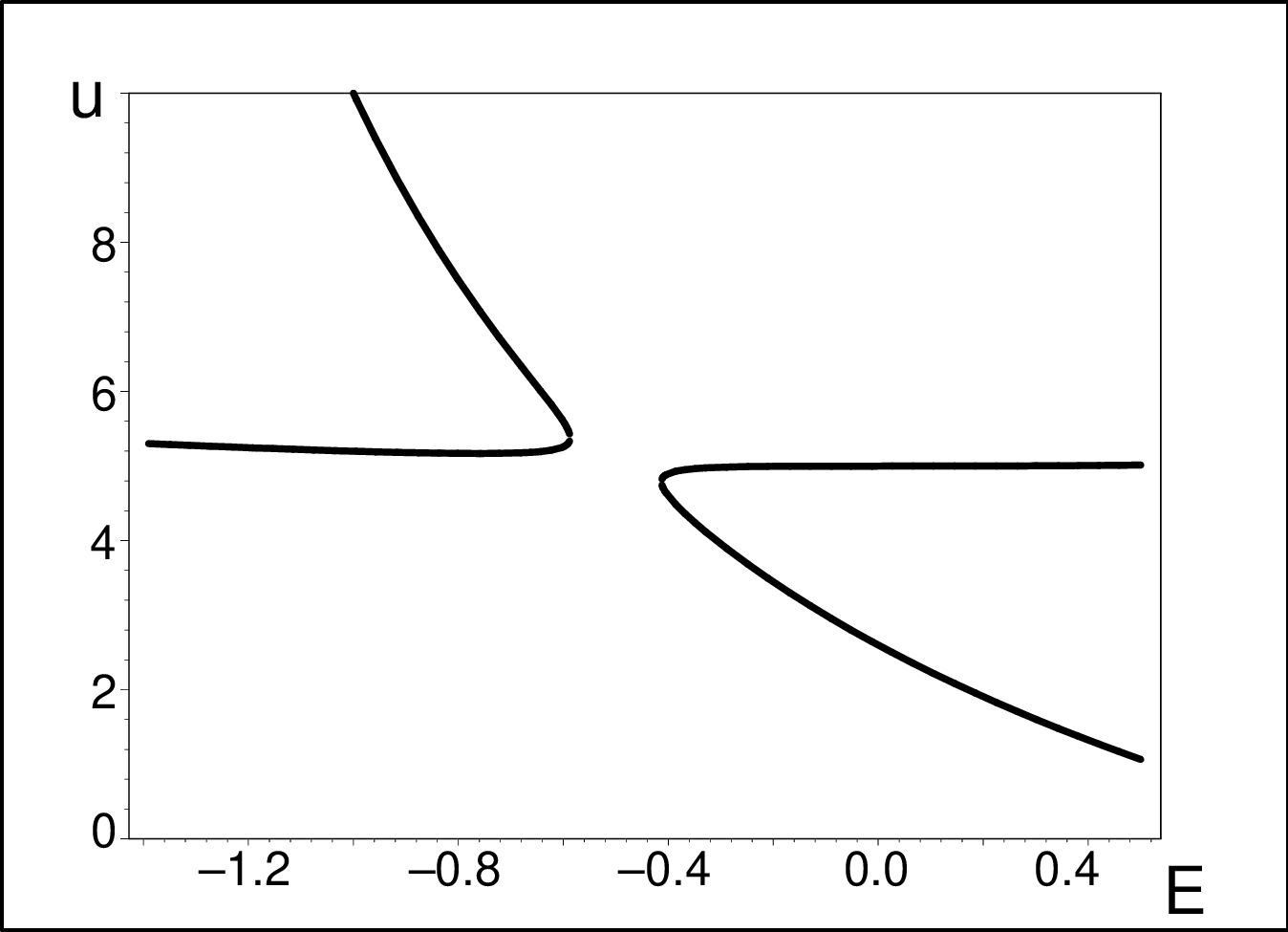,angle=270,width=0.35\textwidth}
\end{center}    
\caption{Avoided crossing of energy levels in the
central part of Figure~\ref{Qwer5s}
(magnified).
 \label{Qwetr5a}}
\end{figure}

According to the picture, all of the three energies seem to
be real to the right from the crossing point, i.e.,
at all $u>5$.
To the left
(i.e., more precisely,
in a finite interval of $u\in (-x,5)$ with, roughly, $x \approx 12$),
two of them become
complex, giving rise
to a pair of
resonant states.
Incidentally, the latter tentative
hypothesis was immediately
disproved
when we magnified the snapshot.
The result is displayed in Figure \ref{Qwetr5a}.

We interchanged the axes $E$ and $u$
there
because this simplifies the analysis
as well as the discussion.
The point is that in Eq.~(\ref{toysek}),
our secular polynomial is cubic in $E$ but merely
quadratic in $u$.
The explicit formulae for
the inverse-function bound-state solutions $u(E)$ are
perceivably simpler, therefore.

In the light of the new, magnified picture,
it becomes obvious that
in contrast to the impression as
evoked, possibly, by the less detailed Figure \ref{Qwer5s},
the
real-level crossing
is safely avoided.
Near the expected point of crossing,
the two real branches of $u(E)$
remain well separated.
More precisely,
both of them cease to be purely real
inside a safely non-empty
interval of $E$ with
slightly negative boundaries.


\subsection{Difficulties with the numerical localization of the true crossing}

The above-mentioned replacement of the energy eigenvalues $E(u)$
by what is called, sometimes, Sturmian eigenvalues $u(E)$ \cite{Sturm}
enabled us to
simplify the discussion and, in particular, to
deduce that
there are just two inverse-function solutions $u_\pm(E)$.
After our choice of
$u^{(EP3)}=5$ they may be shown to
coincide at
$E_-=-0.5868951400$
and at
$E_+=-0.4131048600$ in a way which
is well visible in Figure \ref{Qwetr5a}.

As we already mentioned above, the latter, auxiliary
inverse functions of the continuously variable energy
remain,
strictly inside the open interval
of variable
$E \in (E_-,E_+)$, manifestly non-real
and mutually complex conjugate.
Out of this interval
they are real and non-intersecting,
with an irrelevant polar or
removable singularity at $E=-2$
(can be also seen in Figure \ref{Qwer5s}).

As we already marginally indicated,
their use also simplifies the
graphical presentation of the real bound-state part of the
spectrum, reconfirming
the purely algebraic result
that
equation $u(E)=5$ has merely a single
(viz., triply degenerate) root $E=E^{(EP3)}=0$.

The existence of such a correct EP3-representing root
is also indicated, but not too easily seen,
in the graph of Figure \ref{Qwetr5a}. Still,
its nature becomes more clear
after one returns to algebraic formulae.
In their light, first of all,
we managed to localize
the three-fold EP3 degeneracy instant exactly.

In its small
vicinity,
secondly,
our knowledge of the two Sturmians
enabled us to deduce, for the larger one, the approximation
 \be
  u(E) =5+{ {5}}{{\it E}}^{3}/{24}+O \left( {{\it E}}^{4} \right)\,.
  \label{[33]}
  \ee
It clearly implies that
for $E \approx E^{(EP3)}=0$
(i.e., near the EP3 singularity)
the
three-level $u-$parametrized spectrum
$\{E_j(u)\}$
 unfolds, at $u\neq u^{(EP3)}$,
  into the
 two complex branches $E_{1,2}$
 (resonances, not displayed in our real-spectra-representing pictures)
 and into a single real-energy branch such that
  $$
  E_3(u) \sim (u-u^{(EP3)})^{1/3}\,.
  $$
Unfortunately,
it is still difficult to see
the shape of
the latter, bound-state-representing
branch of energy near  $u^{(EP3)}=5$
in the scale used in Figure \ref{Qwetr5a}.

\begin{figure}[h]                    
\begin{center}                         
\epsfig{file=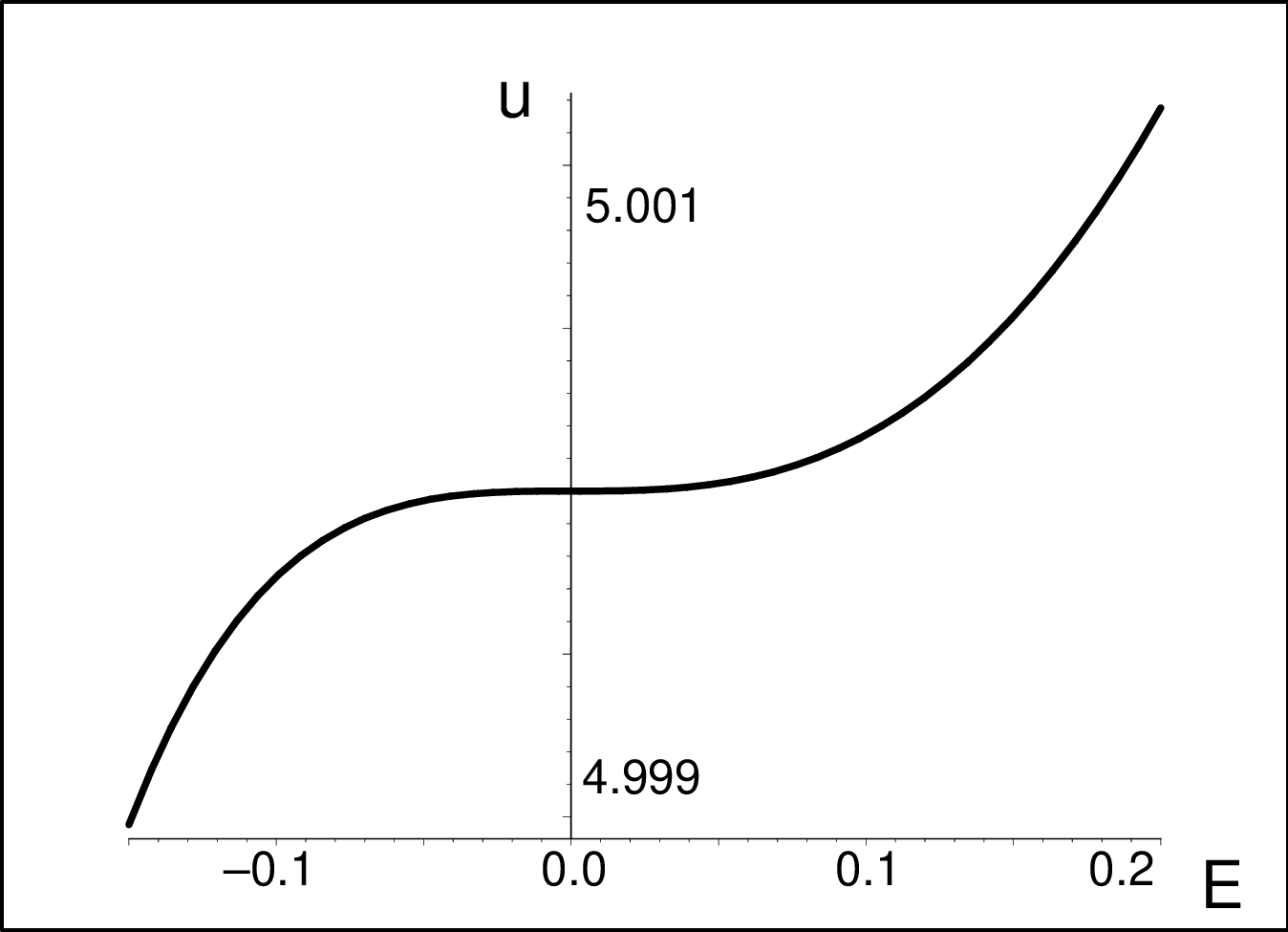,angle=270,width=0.35\textwidth}
\end{center}    
\caption{Magnified central part of Figure \ref{Qwetr5a}.
 \label{Qwetr5b}}
\end{figure}

Fortunately, the availability of the
algebraic form of both of the Sturmians
renders another magnification easy, yielding
our last  Figure \ref{Qwetr5b}. This Figure provides, indeed,
the most detailed
display of the shape (\ref{[33]}) of the upper branch of
Sturmian $u(E)$ near the
innocent-looking true physical
EP3 singularity localized, precisely, at $E=0$.

\newpage

\section{Summary\label{summary}}

A not yet fully available transfer of the theory
of evolution singularities
from its well known
classical version
called ``catastrophe theory''
(attributed, usually, to Ren\'{e} Thom \cite{Zeeman})
to its quantum reincarnation
would be, for different reasons, difficult.
In the QHQM framework,
a visible progress has been achieved, recently,
mainly due to several
truly innovative combinations
of the physics and mathematics behind
a broad class of the energy-level EP
(i.e., exceptional-point) degeneracies.

The roots of this direction of research can be traced back
to the older, quantum-field-theory-motivated
developments in perturbation theory \cite{Kato}.
The studies
of the applicability of these results
in physics
involved, typically, just various
schematic one-dimensional
quantum-mechanical
oscillators admitting, exclusively,
just the
most elementary (i.e.,
two-level,
EP2) degeneracies.
Emerging, moreover, at the mere
complex,
i.e., manifestly unphysical
coupling constants (cf., e.g., papers
\cite{BW,Alvarez}).
Thus, one could conclude that
in this context, the authors
were almost exclusively
motivated by the
underlying mathematics, i.e., e.g.,
by the questions of convergence \cite{Kato}.

The situation has changed when
Bender with Boettcher \cite{BB}
pointed out that the complex
coupling constants
need not be interpreted as unphysical.
Attention has been turned again to the
related quantum EP singularities
and, in particular,
to their possible new role in quantum phenomenology
(cf., e.g., \cite{Berry,Heiss}).

In this context we reopened, in
our present paper,
the problem of the methodically challenging
enormous increase of the technical difficulties encountered
during
multiple attempts of transition from the models with EP2
(as sampled, e.g., in paper \cite{Bijan} -- see also
paragraph \ref{tootsie} above)
to the models with EPN at any higher $N \geq 3$
(such models seem to be truly
exceptional -- cf., e.g., \cite{passage}).
In
this sense we succeeded.
We managed to pay a very constructive attention
to the  energy-level degeneracies
emerging in a specific Swanson-like three-site fermionic
quantum system. For this system we were able to
prove the existence and to localize, first of all,
its Kato's EP=EP3 singular extreme.

\textcolor{black}{Needless to add that an important
source of our success was} the (not quite obvious) exact solvability
of our new model.
\textcolor{black}{We must admit that in a broader physical context
and, in particular, in the context of the
most recent EP-related research
(cf., e.g., its samples in papers \cite{Rotter,EPsa,EP3}),
such a ``user-friendliness'' feature
of the realistic models is rather exceptional.
Even in our models, it was not even initially expected
\cite{Bijan}. The more it helped us}
to
study some of the energy-level-degeneracy-related
characteristics
\textcolor{black}{of our model}
in the vicinity
of its singularity.

Along these lines we found
our model
tractable
by both the graphical and
closed algebraic mathematical means.
In the context of physics we showed,
last but not least, that inside
certain explicitly specified boundaries of the
domain of ``admissible'' coupling constants,
the system
can be considered unitary
and possessing a more or less entirely
conventional
(i.e., quasi-Hermitian \cite{book})
probabilistic interpretation.



\newpage

\subsection*{Funding:}

The project did not receive any support.

\subsection*{Data availability statement:}

No new data were created or analyzed in this study.

\subsection*{Conflicts of Interest:}

The authors declare no conflicts of interests.

\subsection*{ORCID ID:}

  \noindent
Bijan Bagchi: https://orcid.org/0000-0002-0537-2423

  \noindent
Aritra Ghosh: https://orcid.org/0000-0003-0338-2801

  \noindent
Miloslav Znojil: https://orcid.org/0000-0001-6076-0093

\subsection*{Acknowledgements}

A. G. is grateful to Akash Sinha for related discussions.

\end{document}